\documentclass[11pt, letterpaper]{article}

\usepackage[left=1in, right=1in, top=1in, bottom=1in]{geometry}
\usepackage[utf8]{inputenc}
\usepackage[T1]{fontenc}
\usepackage{bm}
\usepackage{type1cm}
\usepackage{lettrine}
\usepackage{amsthm}
\usepackage{moreverb}
\usepackage{mathtools}
\usepackage{amsmath}
\usepackage{amssymb}
\usepackage{algorithmic}
\usepackage{graphics}
\usepackage{graphicx}
\usepackage{subfigure}
\usepackage{caption}
\usepackage{extarrows}
\usepackage{color}
\usepackage{framed}
\usepackage{wrapfig}
\usepackage{bm}
\usepackage{mathrsfs}
\usepackage{mathabx}
\usepackage{multirow}
\usepackage{longtable}
\usepackage{hyperref}
\usepackage{paralist}
\usepackage{indentfirst}
\usepackage{relsize}
\usepackage{upgreek}
\usepackage{mwe}
\usepackage[dvipsnames,table]{xcolor}
\usepackage{booktabs}
\usepackage{authblk}
\usepackage{threeparttable}
\usepackage{placeins}
\usepackage{makecell}
\usepackage{overpic}
\usepackage[sort&compress,numbers]{natbib}
\usepackage[figurename=Figure]{caption}
\usepackage[normalem]{ulem}

\graphicspath{ {./Figure/} }
\usepackage[font=footnotesize,labelfont=bf]{caption}

\providecommand{\keywords}[1]{\textbf{\textit{Keywords: }} #1}

\definecolor{FuPurple}{RGB}{128,0,160}

\hypersetup{
bookmarksopen=true,
bookmarksnumbered=true,
unicode=false,
pdftoolbar=true,
pdfmenubar=true,
pdffitwindow=false,
pdfstartview={FitH},
pdftitle={My title},
pdfauthor={Author},
pdfsubject={Subject},
pdfcreator={Creator},
pdfproducer={Producer},
pdfkeywords={keywords},
pdfnewwindow=true,
colorlinks=true,
linkcolor=blue,
citecolor=blue,
filecolor=blue,
urlcolor=blue
}

\begin{document}

\title{\textbf{Scaling Neural Network Quantum States for\\ \textit{Ab Initio} Quantum Chemistry}}

\author[1,2,$^\dag$]{Chenxi Yu}
\author[1,2,$^\dag$]{Hanlin Kong}
\author[1,2,$^\dag$]{Jianan Wei}
\author[3,$^\dag$]{Lizhong Fu}
\author[3,$^*$]{\\Honghui Shang}
\author[1,2,$^*$]{Wenguan Wang}
\author[3]{Jinlong Yang}

\affil[1]{\small State Key Lab of Brain-Machine Intelligence, Zhejiang University}
\affil[2]{\small College of Artificial Intelligence, Zhejiang University}
\affil[3]{\small State Key Laboratory of Precision and Intelligent Chemistry, University of Science and Technology of China \vspace{18pt}} 

\affil[*]{Corresponding authors}
\affil[$\dag$]{Equal contribution}

\date{}

\maketitle

\normalsize

\vspace{-18pt} 
\begin{abstract}
	\small
    Neural-network quantum states (NNQSs) can represent many-electron wave functions without explicitly enumerating the determinant space, but their accuracy depends jointly on model size and variational-optimization effort. 
    Here we characterize this dependence for a physics-conditioned autoregressive NNQS trained separately on two six-molecule source benchmarks. 
    Across eight model sizes and five optimization milestones, we find that model size and optimization steps jointly shape the energy error. The capacity advantage of larger models becomes more apparent with sufficient optimization, while the returns from additional optimization vary with model size. 
    We capture this coupling using an interaction scaling law and quantify the cumulative compute of each evaluated configuration. 
    The resulting error--compute Pareto frontiers provide a practical decision rule for jointly selecting model size and optimization steps under a given compute budget within the evaluated range.
    Furthermore, we find that this beneficial scaling trend persists during fine-tuning on held-out N$_2$. Pretrained models show decreasing error with increasing model size, with a steeper reduction following pretraining on the Hard benchmark.
    Together, these results place autoregressive neural quantum states within the broader landscape of empirical neural scaling and open a quantitative route toward the systematic scaling of neural quantum solvers for \textit{ab initio} quantum chemistry.
\end{abstract}


\vspace{12pt} 
\section{Introduction}

Large language models (LLMs) based on autoregressive Transformers have demonstrated remarkable capabilities across a wide range of natural language tasks~\cite{Vaswani2017Attention,Brown2020Language,thoppilan2022lamda}. 
Their rapid development has been driven by the continued growth of model parameters, training data, and computational resources, together with an improved understanding of the predictable relationships between these factors and model performance~\cite{Hestness2017Deep,Kaplan2020Scaling}. 
Previous studies have shown that language-model loss approximately follows power-law relationships as model size, dataset size, and computational budget increase; these empirical relationships are commonly referred to as scaling laws~\cite{Kaplan2020Scaling,Henighan2020Scaling,rae2021scaling,Hoffmann2022Training,Alabdulmohsin2022Revisiting}. 
Scaling laws allow researchers to predict the potential gains from further scaling and to allocate resources among model size, data volume and computational cost in a principled manner. As the core architecture of modern language models, the autoregressive Transformer decomposes a complex joint distribution into a sequence of conditional distributions, providing a unified framework for large-scale sequence modeling. Understanding the scaling behavior of autoregressive Transformers is therefore important for revealing how their capabilities grow, identifying performance bottlenecks and guiding the efficient development of models~\cite{bahri2024explaining}.

Beyond language modeling, the combination of expressive neural representations and stochastic optimization has also opened new directions for solving quantum many-body problems. For molecular systems, the computational objective is determined by first-principles quantum mechanics rather than by fitting externally provided labels~\cite{Hermann2023initio}. Within the Born--Oppenheimer approximation and a finite orbital basis, the nuclear charges and coordinates define the electronic Hamiltonian, while the electron number specifies the corresponding many-electron Hilbert space~\cite{McArdle2020Quantum}. The electronic wave function can be represented as a superposition of occupation-number configurations, whose number increases combinatorially with the number of orbitals and electrons, making direct enumeration of the full Hilbert space rapidly infeasible~\cite{Kim2021Flexible}. Variational Monte Carlo (VMC) overcomes this limitation by optimizing parameterized wave functions using stochastic estimates of the energy and its gradients~\cite{Foulkes2001Quantum}. 
Neural-network quantum states (NNQSs) further extend VMC by introducing neural networks as expressive and differentiable wave-function ansatzes, enabling flexible representations of complex many-body states and high-accuracy electronic-structure calculations~\cite{Carleo2017Solving,Choo2020Fermionic,Hermann2020Deepneuralnetwork,Lange2024architectures}.
Unlike conventional supervised learning models that minimize prediction errors against labeled data, NNQSs are optimized by minimizing the variational energy of the underlying Hamiltonian. Their achievable accuracy is therefore governed by the interplay among neural-network expressivity, stochastic sampling efficiency and optimization dynamics, raising a fundamental question of how these factors determine performance scaling as NNQS models and computational resources grow.

Autoregressive NNQSs provide a natural framework for studying such scaling behavior by connecting neural quantum states with modern generative modeling. By decomposing the probability distribution of many-electron configurations into a sequence of normalized conditional probabilities, autoregressive models enable direct sampling of electronic configurations without the burn-in, accept--reject steps and autocorrelation effects associated with Markov-chain Monte Carlo methods~\cite{Sharir2020Deep,HibatAllah2020RNN,Barrett2022Autoregressive}. This property is particularly valuable for molecular electronic structure calculations, where the determinant distribution can be highly non-uniform and the evaluation of Hamiltonian-connected configurations constitutes a major computational cost~\cite{Barrett2022Autoregressive,Zhao2023Scalable}. Recent Transformer-based quantum states further improve the representation of long-range dependencies between orbital occupations through causal self-attention mechanisms~\cite{Zhang2023Transformer,Viteritti2023Transformer,Sprague2024Variational}. These advances have enabled highly accurate molecular ground-state calculations using autoregressive neural quantum states, including unified architectures that jointly model wave-function amplitudes and phases~\cite{Shang2025Solving}. Existing studies~\cite{Jiang2025NeuralScaling,Knitter2025Scaling,Rende2026Scaling,Lu2026InformationTheoretic} have investigated model-size and compute scaling in molecule-specific neural wave functions, while the coupled scaling of model size and optimization in a shared, physics-conditioned autoregressive NNQS jointly optimized across multiple molecules remains largely unexplored. A systematic understanding of how model size and optimization progress jointly determine accuracy under a fixed computational budget remains an open challenge.

Against this background, we investigate scaling in a unified molecular setting, where a single physics-conditioned autoregressive NNQS is optimized jointly across multiple molecules. Our primary question is whether model size and optimization progress exert separable effects on energy accuracy, or whether the benefit of increasing one depends on the level of the other. A larger model may provide greater expressive capacity, yet this capacity advantage may not be fully realized at early stages of optimization. Conversely, the marginal benefit of additional optimization may vary across model sizes. Characterizing this model-size--optimization interaction can therefore reveal how representational capacity and optimization progress jointly shape wave-function accuracy, distinguish capacity-limited from optimization-limited regimes, and determine when additional model capacity becomes effective. We further ask whether such scale-dependent behavior persists when the learned representation is adapted to a Hamiltonian not encountered during source training.

We study this question using QiankunNet~\cite{Shang2025Solving}, an autoregressive Transformer-based NNQS, for multi-molecule \textit{ab initio} quantum chemistry. 
The physics-conditioned autoregressive NNQS comprises a geometry encoder and a Hamiltonian encoder that integrate molecular geometry and Hamiltonian information into a molecule-specific condition, allowing one model to be optimized across multiple molecular systems.
Building on this conditional architecture, we construct a family of eight model sizes and train separate instances of each size on the Easy and Hard multi-molecule benchmarks. 
We evaluate the per-molecule mean absolute energy error at predefined optimization milestones and characterize its joint dependence on model size and optimization steps using an interaction power law. 
Separately, to translate this accuracy landscape into a resource-allocation criterion, we estimate the cumulative computational cost of each evaluated configuration and construct empirical error--compute Pareto frontiers. These frontiers identify the model size and optimization steps that achieve the lowest measured error under a given compute budget. 
Finally, we test whether the model-size advantage established during pretraining transfers to a held-out N$_2$ Hamiltonian. Fine-tuning on N$_2$ shows that pretrained models retain a favorable model-size trend, whereas randomly initialized models do not exhibit a comparable trend under the same fine-tuning budget.

\section{Results}

\subsection{Coupled scaling with model size and optimization steps}

\begin{figure}[t]
  \centering
  \includegraphics[width=0.85\textwidth]{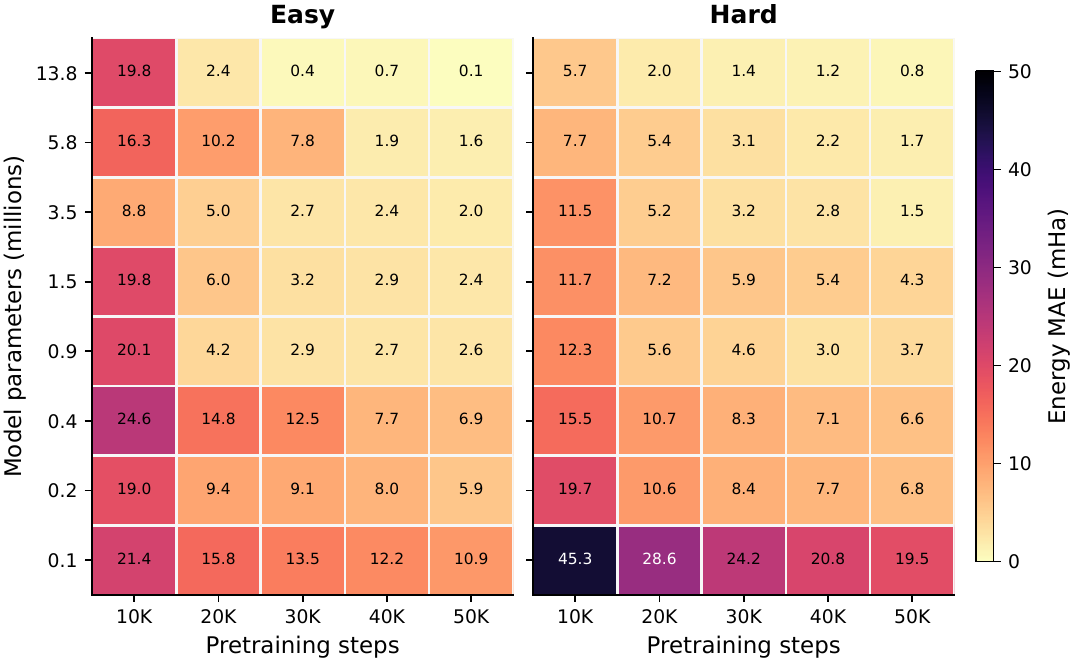}
  \caption{
    \textbf{Energy error across model sizes and optimization milestones.}
    Mean absolute energy error (MAE), reported in millihartrees (mHa), is evaluated across eight model sizes and five cumulative optimization milestones for the Easy (left) and Hard (right) benchmarks.
    Rows denote the number of trainable parameters in millions, and columns denote cumulative optimization steps. 
    Each cell reports the equally weighted mean MAE over molecules, taken over the 1,000-update window ending at the corresponding milestone, with the error magnitude indicated by both color and annotated values.
  }
  \label{fig:joint-error-surface}
\end{figure}

We first examine how model size and optimization steps jointly shape energy accuracy during multi-molecule pretraining. 
Based on the size of the full configuration interaction (FCI) space, we construct two molecular benchmarks of increasing complexity, denoted Easy and Hard (Table~\ref{tab:molecular-benchmarks}; Methods Section~\ref{sec:methods-multimolecule-transfer}; Supplementary Section~\ref{sec:supp-molecular-systems}). 
We consider eight model sizes by varying the hidden dimension and the number of Transformer blocks, spanning approximately 0.1M to 13.8M trainable parameters. 
Each model is optimized for up to 50K steps, with energy errors evaluated at 10K-step milestones.

Across both benchmarks, the mean absolute energy error (MAE) generally decreases with continued optimization at a fixed model size, although individual checkpoints exhibit non-monotonic fluctuations associated with stochastic VMC optimization (Fig.~\ref{fig:joint-error-surface}). 
The marginal improvement from additional optimization tends to diminish as training proceeds, but the onset of this saturation depends on benchmark complexity. 
Models on the Easy benchmark approach a plateau at earlier stages, whereas those on the Hard benchmark continue to benefit from additional optimization over a longer range of steps. 
Within the evaluated range, the Hard benchmark therefore requires a larger optimization budget before the gains begin to saturate.

The dependence on model size shows a complementary pattern. 
At fixed optimization steps, the benchmark-level trend generally favors larger models, although substantial transient non-monotonicity is observed, particularly at early optimization stages (Supplementary Fig.~\ref{fig:supp-benchmark-parameter-scaling}). 
Molecule-resolved results nevertheless exhibit substantial heterogeneity and transient non-monotonic behavior, particularly on the Easy benchmark (Supplementary Fig.~\ref{fig:supp-molecule}). 
On the Easy benchmark, for example, the 3.5M-parameter model achieves the lowest error at 10K steps, while the larger 5.8M- and 13.8M-parameter models have not yet realized their capacity advantage. 
This delayed benefit remains evident at 20K--30K steps, where the 5.8M-parameter model is still outperformed by several smaller configurations. 
With continued optimization, these ranking inversions are progressively reduced, and by 50K steps the benchmark-level dependence becomes much more consistently aligned with model size. 
These results indicate that the accuracy advantage of increased capacity emerges most clearly when accompanied by sufficient optimization.

These observations suggest that a separable description of model size and optimization steps may be insufficient. 
We therefore fit the complete model-size--optimization grid using separable and interaction power laws, together with a more flexible quadratic log-response surface (Supplementary Section~\ref{sec:supp-joint-scaling}). 
The interaction power law is preferred by AICc on both benchmarks. 
Relative to the separable model, its AICc is lower on both Easy ($-49.70$ versus $-35.09$) and Hard ($-102.36$ versus $-96.14$), and it is also favored over the quadratic response surface by AICc despite using fewer coefficients (Supplementary Table~\ref{tab:joint-model-selection}). 
The resulting joint scaling relation is:
\begin{equation}
\log\left[\frac{\mathcal{L}(P,S)}{1\,\mathrm{mHa}}\right]
=
\log\left[\frac{L_0}{1\,\mathrm{mHa}}\right]
-a\log\left(\frac{P}{P_{\mathrm{ref}}}\right)
-b\log\left(\frac{S}{S_{\mathrm{ref}}}\right)
-\eta
\log\left(\frac{P}{P_{\mathrm{ref}}}\right)
\log\left(\frac{S}{S_{\mathrm{ref}}}\right),
\label{eq:joint-interaction-scaling}
\end{equation}
where \(P\) is the number of trainable parameters, \(S\) is the cumulative number of optimization steps, \(L_0\) is the fitted reference error, and \(P_{\mathrm{ref}}=10^{6}\) and \(S_{\mathrm{ref}}=10^{4}\) are fixed reference scales. 
The coefficient \(\eta\) quantifies the coupling between model size and optimization steps. 
The fitted interactions are positive on both benchmarks, with \(\eta=0.378\) for Easy and \(\eta=0.133\) for Hard (Supplementary Table~\ref{tab:joint-fit-coefficients}). 
Within the evaluated range, a positive interaction indicates that the error reduction associated with increasing model size becomes stronger as more optimization is provided, quantitatively capturing the delayed emergence of the large-model advantage observed above.

\begin{figure}[t]
  \centering
  \includegraphics[width=0.85\textwidth]{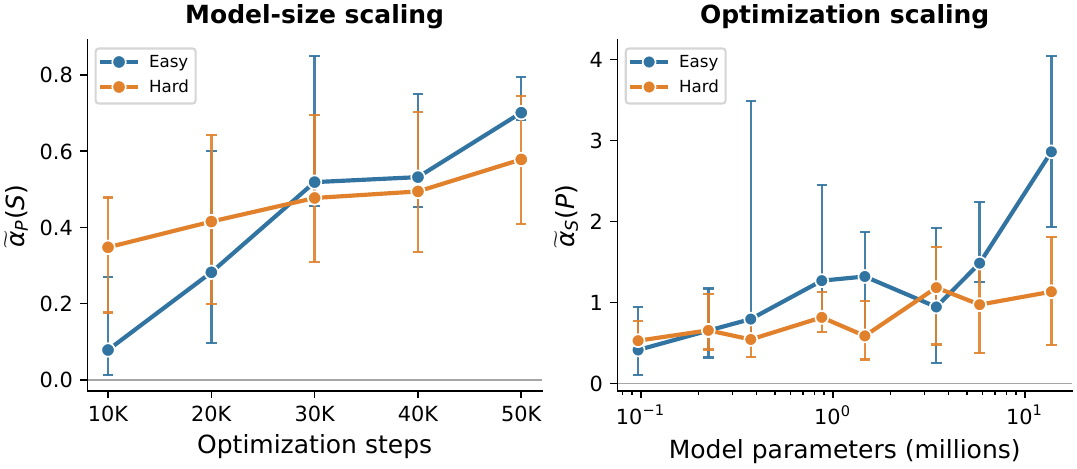}
  \caption{
    \textbf{Effective scaling exponents as functions of optimization steps and model size.}
    For the Easy and Hard benchmarks, \(\widetilde{\alpha}_{P}(S)\) is fitted across eight model sizes at each fixed optimization milestone (left), whereas \(\widetilde{\alpha}_{S}(P)\) is fitted across five optimization milestones at each fixed model size (right). The bars indicate the range of fitted exponents across individual molecules within each benchmark. Larger positive exponents indicate a faster reduction in MAE as the corresponding resource increases.
  }
  \label{fig:conditional-exponents}
\end{figure}

To examine the same behavior independently of the joint interaction fit, we further perform separate one-axis power-law fits along fixed slices of the model-size--optimization grid (Fig.~\ref{fig:conditional-exponents}; Supplementary Section~\ref{sec:supp-joint-scaling}). 
At each optimization milestone, the model-size exponent \(\widetilde{\alpha}_{P}(S)\) is fitted across the eight model sizes, whereas at each model size, the optimization-step exponent \(\widetilde{\alpha}_{S}(P)\) is fitted across the five milestones. 
As optimization increases from 10K to 50K steps, \(\widetilde{\alpha}_{P}(S)\) rises from 0.079 to 0.703 on Easy and from 0.349 to 0.580 on Hard. 
This progressively steeper model-size dependence provides consistent empirical evidence that additional capacity becomes more effective with continued optimization. 
By contrast, \(\widetilde{\alpha}_{S}(P)\) varies non-monotonically across model sizes, reaching 2.859 on Easy and 1.131 on Hard at the largest model size. 
Thus, the clearest and most consistent manifestation of the coupling is the strengthening of the model-size advantage with optimization, whereas the marginal return from additional optimization exhibits greater scale-dependent variability.

\begin{figure}[!t]
  \centering
  \includegraphics[width=0.7\textwidth]{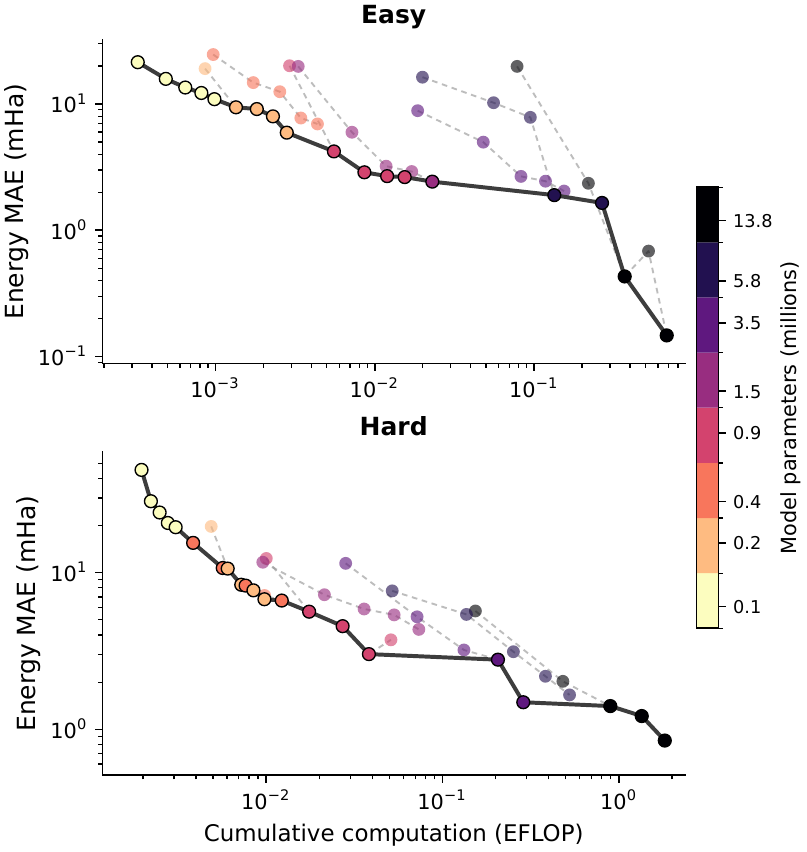}
  \caption{
    \textbf{Error--compute Pareto frontiers over the evaluated grid of model sizes and optimization milestones.} MAE (mHa) is plotted against cumulative compute (EFLOP) for the Easy (top) and Hard (bottom) benchmarks. Each point represents one evaluated combination of model size and optimization milestone, with color indicating the number of trainable parameters. Dashed lines connect milestones from the same model size, while solid black lines trace the empirical Pareto frontiers formed by non-dominated configurations. 
  }
  \label{fig:compute-pareto}
\end{figure}

\begin{figure}[!t]
  \centering
  \includegraphics[width=\textwidth]{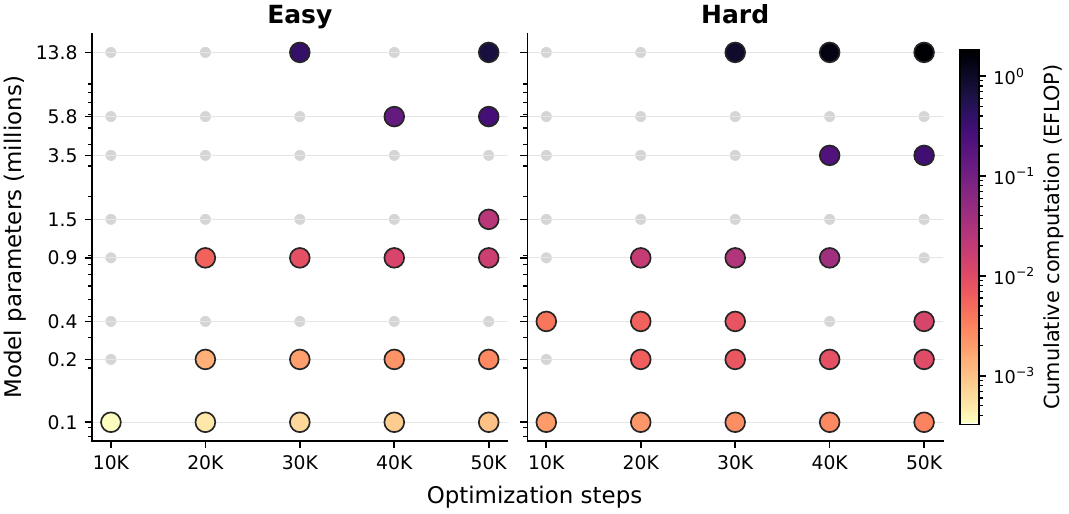}
  \caption{
    \textbf{Pareto-optimal configurations over the evaluated grid of model sizes and optimization milestones.}
    Each panel shows a grid of model size versus optimization milestone for the Easy (left) and Hard (right) benchmarks. 
    Rows denote the number of trainable parameters in millions, and columns denote cumulative optimization steps. 
    Grey circles indicate dominated configurations, whereas colored circles with black outlines indicate the non-dominated configurations. 
    Color encodes cumulative compute (EFLOP) on a logarithmic scale.
  }
  \label{fig:compute-optimal-configurations}
\end{figure}

\subsection{Empirical compute-optimal allocation of model size and optimization steps}

The observed coupling between model size and optimization steps raises a practical question: how should these two resources be allocated under a finite compute budget? We address this question by estimating the cumulative computational cost of each evaluated model--milestone configuration and examining the resulting error--compute trade-off. We define the empirical Pareto frontier as the set of non-dominated configurations for which no other evaluated configuration achieves both lower computational cost and lower energy error.
The compute accounting and frontier construction procedures are described in Methods Sections~\ref{sec:methods-model-compute} and~\ref{sec:methods-scaling-analysis} and Supplementary Section~\ref{sec:supp-compute-accounting}.

Across both benchmarks, larger compute budgets allow progressively lower minimum errors within the evaluated configuration grid, but the corresponding Pareto-optimal allocation between model size and optimization steps is not fixed. 
At relatively low compute, the frontier is primarily composed of smaller models evaluated at earlier optimization milestones. 
As the compute budget increases, additional optimization of an existing model can remain more efficient than immediately increasing model size. 
Larger models therefore become Pareto-optimal only after sufficient compute is available to optimize them effectively. 
This progression is consistent with the model size--optimization coupling observed above: additional capacity provides an accuracy advantage only when accompanied by an adequate optimization budget.

The detailed frontier structure depends on benchmark complexity. On the Easy benchmark, the Pareto-optimal points are well described, over the evaluated compute range, by a single power-law relationship with a fitted exponent of 0.512. The Hard benchmark instead shows two empirical compute regimes, with fitted exponents of 1.062 and 0.367, accompanied by shifts in the Pareto-optimal model size--optimization-step combinations (Figs.~\ref{fig:compute-pareto} and~\ref{fig:compute-optimal-configurations}; Supplementary Section~\ref{sec:supp-pareto-fitting}). Thus, the configuration that uses compute most efficiently depends on both the available budget and the complexity of the molecular benchmark. Within the regime studied here, these results support joint selection of model size and optimization steps rather than a fixed allocation rule across compute budgets.

\subsection{Transfer of model-size scaling to a held-out \texorpdfstring{N$_2$}{N2} Hamiltonian}
\label{results:transfer_N2}

\begin{figure}[!th]
  \centering
  \includegraphics[width=0.55\textwidth]{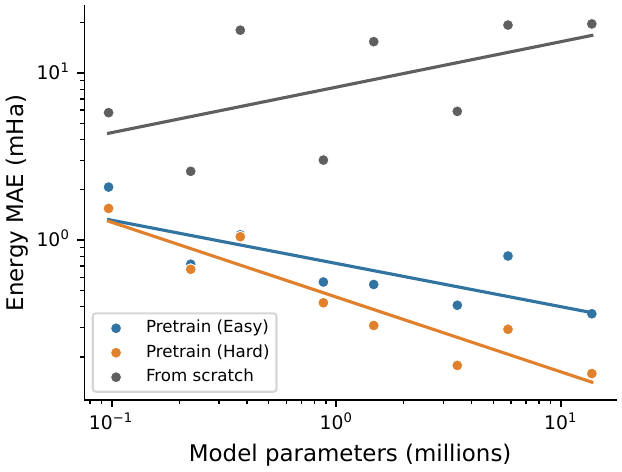}
  \caption{
    \textbf{N$_2$ fine-tuning error across model sizes and initialization strategies.}
    Energy MAE (mHa) after fine-tuning on N$_2$ is shown across eight model sizes for models initialized from Easy pretraining (blue), Hard pretraining (orange), and random weights (grey). 
    Solid lines denote descriptive power-law fits in log--log space.
  }
  \label{fig:n2-transfer}
\end{figure}

Having established the joint dependence of pretraining accuracy on model size and optimization steps, we next examine whether the model-size advantage extends to fine-tuning on an unseen Hamiltonian. We fine-tune the same eight model sizes on N$_2$, using Easy-pretrained, Hard-pretrained, or randomly initialized weights for the same number of fine-tuning steps, and evaluate the resulting energy errors as a function of model size.

The two pretrained series retain a clear model-size dependence during fine-tuning, whereas the randomly initialized series does not (Fig.~\ref{fig:n2-transfer}). 
For both Easy and Hard pretraining, the energy error decreases approximately as a power law with increasing model size, with fitted exponents of 0.259 and 0.447, respectively. By contrast, randomly initialized models show no clear decrease in error with increasing model size under the fixed fine-tuning budget, with a fitted exponent of -0.272.

The absence of a favorable model-size trend under random initialization is consistent with the scale-dependent optimization behavior observed during pretraining. Under a fixed and limited fine-tuning budget, larger randomly initialized models require more optimization to make effective use of their additional capacity and can therefore remain underoptimized as model size increases. Pretrained models start from parameters already adapted to molecular electronic structure, reducing this optimization burden and allowing the advantage of larger models to emerge within the same number of fine-tuning steps. Thus, pretraining not only improves the absolute fine-tuning accuracy but also preserves the favorable dependence of accuracy on model size.

Among the two pretrained initializations, Hard pretraining yields the stronger size-dependent reduction in error and achieves the lowest error across all evaluated model sizes. At the largest model size, the errors are 0.363~mHa for Easy pretraining and 0.159~mHa for Hard pretraining, compared with 19.67~mHa for random initialization. The fitted scaling parameters and descriptive fit statistics are reported in Supplementary Section~\ref{sec:supp-n2-transfer}.

\section{Discussion}

In this study, we investigate the scaling behavior of autoregressive NNQSs for molecular electronic structure using a scalable model family, multi-molecule benchmarks of different complexities, and a unified evaluation protocol. Across eight model sizes and five optimization milestones, we characterize the joint dependence of energy accuracy on model size and optimization steps, construct empirical error--compute Pareto frontiers, and examine whether the model-size trend persists during fine-tuning to N$_2$.

We find that model size and optimization steps exhibit coupled empirical scaling. Across both benchmarks, the model-size scaling exponent increases with continued optimization, indicating that the advantage of larger models becomes progressively clearer as more optimization is provided. We capture this dependence using a power-law model with an interaction term (Eq.~\eqref{eq:joint-interaction-scaling}; Supplementary Table S4). The slice-wise fits in Fig. 2 provide an independent empirical view of the same coupling. Similar to the resource-allocation perspective of Chinchilla-style scaling~\cite{Hoffmann2022Training}, these results suggest that the two scaling variables should be considered jointly rather than independently.

One possible explanation is that the models operate in an optimization-limited regime, particularly at early training stages. The transient ranking inversions are consistent with larger models requiring more optimization to realize their capacity advantage. However, the present experiments do not distinguish slower convergence from optimization instability, stochastic VMC fluctuations, or sampling noise. The fitted interaction should therefore be interpreted as a finite-range empirical description rather than an asymptotic law.

This joint scaling is particularly relevant to \textit{ab initio} VMC. Unlike language-model scaling, which is commonly organized around model size and data volume~\cite{Kaplan2020Scaling,Hoffmann2022Training}, the molecular solution in VMC is obtained through iterative optimization. The computational cost therefore depends jointly on model size, optimization steps, and the sampling-dependent workload incurred at each step. Our compute proxy captures this accumulated neural and optimizer workload, while remaining an approximation to the full end-to-end VMC cost.

Under this compute proxy, the Pareto-optimal configuration varies with the available budget and differs between the Easy and Hard benchmarks (Figs.~\ref{fig:compute-pareto} and~\ref{fig:compute-optimal-configurations}). Smaller models are more competitive at low budgets, whereas larger models enter the Pareto frontier only when sufficient optimization becomes affordable. These frontiers therefore provide an empirical criterion for jointly selecting model size and optimization steps under finite computational resources. Their different compositions across the two benchmarks further suggest that the fitted scaling behavior is system dependent.

The transfer experiments further show that the favorable model-size trend established during pretraining can persist during downstream fine-tuning. For models initialized from Easy or Hard pretraining, the N$_2$ error decreases with increasing model size, whereas the same trend is not resolved for training from scratch within the evaluated budget. This contrast is consistent with pretraining allowing larger models to exploit their additional capacity more effectively during limited-budget fine-tuning. The lower errors after Hard pretraining may also reflect differences in benchmark composition or complexity. Because the transfer analysis is restricted to one held-out molecule, these results should not be interpreted as a general transfer scaling law.

Several limitations define the scope of this study. First, our analysis is restricted to autoregressive Transformer-based NNQSs, and whether similar scaling extends to other neural wave-function architectures remains to be established~\cite{Barrett2022Autoregressive,HibatAllah2020RNN}. Second, we consider molecular systems at fixed geometries and do not address geometry-dependent tasks such as potential energy surface modeling~\cite{Gao2022PESNet,Scherbela2022WeightSharing}. Third, the Easy and Hard benchmarks are primarily defined by FCI determinant-space size, so the effects of electronic complexity cannot be fully separated from differences in elemental composition and chemical structure. Broader molecular families, architectures, and geometry-dependent problems will be needed to assess the generality of the observed scaling behavior.

\section{Methods}

\subsection{Preliminaries}

An \textit{ab initio} electronic-structure calculation determines molecular properties from the nuclear charges, nuclear coordinates and electron number by solving the electronic Schrödinger equation within explicit approximations, rather than employing an empirical potential fitted to the molecule of interest.
Under the Born--Oppenheimer approximation and in a chosen finite orbital basis, these physical inputs determine the molecular integrals and hence the electronic Hamiltonian \(H_m\) for molecule \(m\).

Following the standard second-quantization formalism employed in NNQS calculations~\cite{Wu2023NNQSTransformer}, a many-electron configuration in \(N_m\) spin orbitals is represented by an ordered occupation-number bitstring $\bm{x} = (x_1, x_2, \dots, x_{N_m})$, where each entry $x_i$ denotes the occupation of a specific spin orbital.
The total electron number and spin-projection sector dictate the allowed bitstrings. Let \(\mathcal X_m\) denote this fixed-electron-number and fixed-spin-projection sector. The wave function assigns a complex coefficient to each valid configuration:
\begin{equation}
|\Psi_{\theta,m}\rangle
=
\sum_{\bm{x}\in\mathcal X_m}\Psi_{\theta,m}(\bm{x})|\bm{x}\rangle,
\qquad
\Psi_{\theta,m}(\bm{x})
=
|\Psi_{\theta,m}(\bm{x})|e^{\mathrm{i}\phi_{\theta,m}(\bm{x})}.
\label{eq:occupation-wavefunction}
\end{equation}
Here, for a normalized state, \(|\Psi_{\theta,m}(\bm{x})|^2\) is the Born probability of configuration \(\bm{x}\), whereas \(\phi_{\theta,m}(\bm{x})\) represents its sign or phase structure.
In the NNQS framework, a neural network parameterizes both the magnitude and phase of the wave function. 
Specifically, it maps the discrete input bitstring $\bm{x}$ to its corresponding complex coefficient. 
The network weights $\theta$ represent variational degrees of freedom that are optimized to minimize the energy of the Hamiltonian $H_m$. 
By the Rayleigh--Ritz principle~\cite{Foulkes2001Quantum}, the corresponding variational energy satisfies:
\begin{equation}
E_m(\theta)
=\frac{\langle\Psi_{\theta,m}|H_m|\Psi_{\theta,m}\rangle}
{\langle\Psi_{\theta,m}|\Psi_{\theta,m}\rangle}
\geq E_{0,m},
\label{eq:variational-energy}
\end{equation}
where \(E_{0,m}\) is the ground-state energy of \(H_m\) in the target sector and chosen orbital basis.

\subsection{Autoregressive wave-function ansatz}
\label{sec:methods-autoregressive-ansatz}

To parameterize the configuration coefficients in Eq.~\eqref{eq:occupation-wavefunction}, we employ QiankunNet, an autoregressive Transformer-based NNQS with separate amplitude and phase networks~\cite{Shang2025Solving}.
For a molecule \(m\) with \(T_m\) spatial orbitals, the alpha- and beta-spin occupations associated with spatial orbital \(i\) are paired into a four-state token \(v_i\in\{00,10,01,11\}\). 
This converts the spin-orbital bitstring \(\bm{x}\) into the token sequence \(\bm{v}=(v_1,\ldots,v_{T_m})\).
A decoder-only causal Transformer parameterizes the probability of the occupation sequence through the autoregressive factorization:
\begin{equation}
p_{\theta,m}(\bm{v})
=
\prod_{i=1}^{T_m}p_{\theta,m}(v_i\mid v_{<i}),
\label{eq:autoregressive-born-probability}
\end{equation}
whereas a separate multilayer perceptron (MLP) parameterizes the phase \(\phi_{\theta,m}(\bm{v})\) from the complete occupation sequence. 
The resulting complex wave-function coefficient is:
\begin{equation}
\Psi_{\theta,m}(\bm{v})
=
\sqrt{p_{\theta,m}(\bm{v})}\,
\exp\!\left[\mathrm{i}\phi_{\theta,m}(\bm{v})\right],
\label{eq:autoregressive-wavefunction}
\end{equation}
where \(\mathrm{i}=\sqrt{-1}\) denotes the imaginary unit.
Together, Eqs.~\eqref{eq:autoregressive-born-probability}
and~\eqref{eq:autoregressive-wavefunction} separate the two roles of the ansatz: the Transformer represents the Born probability
\(p_{\theta,m}(\bm{v})=|\Psi_{\theta,m}(\bm{v})|^2\), while the phase network \(\phi_{\theta,m}(\bm{v})\) represents the sign or phase relations among configurations. 

\subsection{Variational Monte Carlo}

The variational parameters are optimized by minimizing the energy of each molecular Hamiltonian using variational Monte Carlo (VMC)~\cite{Foulkes2001Quantum}. 
Based on the Born distribution defined in Eq.~\eqref{eq:autoregressive-born-probability}, VMC formulates the molecular energy as the expectation value of the local energy,
\begin{equation}
E_m(\theta)
=
\mathbb{E}_{\bm{v}\sim p_{\theta,m}}
\left[E_{\mathrm{loc},m}(\bm{v})\right].
\label{eq:vmc-energy}
\end{equation}
For a given sampled configuration $\bm{v}$, the local energy is defined as:
\begin{equation}
E_{\mathrm{loc},m}(\bm{v})
=
\sum_{\bm{v}'}
H^{(m)}_{\bm{v}\bm{v}'}
\frac{\Psi_{\theta,m}(\bm{v}')}
{\Psi_{\theta,m}(\bm{v})},
\qquad
H^{(m)}_{\bm{v}\bm{v}'}
=
\langle\bm{v}|H_m|\bm{v}'\rangle.
\label{eq:local-energy}
\end{equation}
Since only configurations $\bm{v}'$ connected to $\bm{v}$ via non-zero Hamiltonian matrix elements contribute to this summation, averaging the local energies over sampled configurations yields an estimate of $E_m(\theta)$ without requiring the full enumeration of the configuration space.
The corresponding stochastic energy gradients are evaluated from these same samples using automatic differentiation and are then used to update the network parameters in each optimization step. 

To evaluate the expectation value in Eq.~\eqref{eq:vmc-energy}, independent configurations must be sampled from the Born distribution.
For this purpose, the sampling is performed using a layer-wise Monte Carlo tree search (MCTS), implemented as batched autoregressive sampling (BAS)~\cite{Sharir2020Deep,Barrett2022Autoregressive,Shang2025Solving}.
By treating the generation of a configuration $\bm{v}$ as a sequential process, this approach maps each level of the sampling tree to one spatial orbital. 
Unlike Markov-chain sampling, BAS generates these orbital occupations sequentially by drawing each token directly from the conditional probability $p_{\theta,m}(v_i\mid v_{<i})$. 
During this process, BAS groups identical partial configurations using integer multiplicities. At each step, these multiplicities are distributed according to the predicted conditional probabilities, while invalid branches are masked to enforce the target alpha- and beta-electron numbers.

\subsection{Physics-conditioned autoregressive NNQS}
\label{sec:methods-physical-conditioning}

\begin{figure}[!ht]
    \centering
    \begin{overpic}[width=\textwidth]{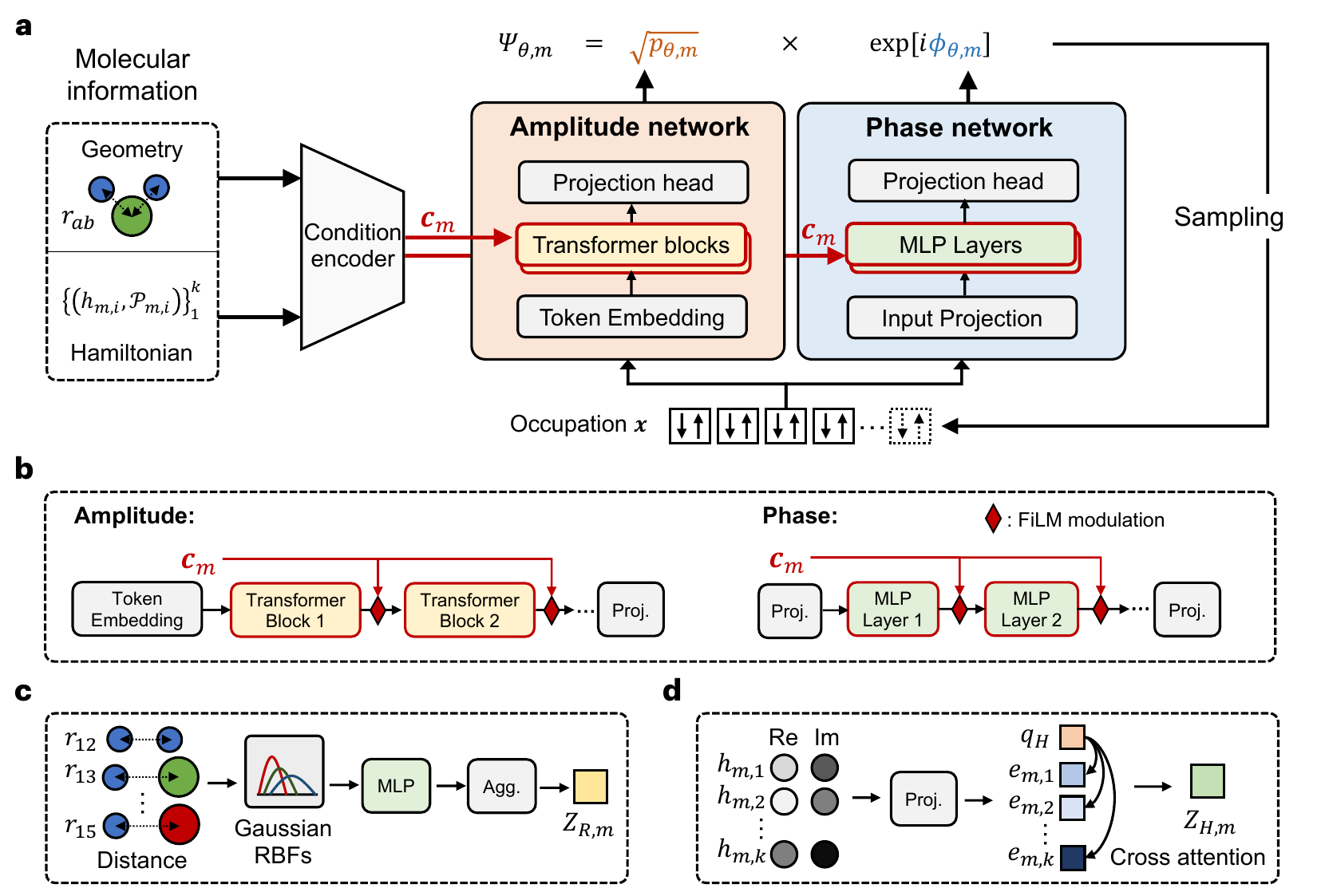}
        \put(89,28){\scriptsize{Eq.~\eqref{eq:physical-film}}}
    \end{overpic}
    \caption{
    \textbf{Physics-conditioned autoregressive NNQS~\cite{Shang2025Solving} for multi-molecule optimization.}
    \textbf{a}, Overview of the shared NNQS architecture.
    Molecular geometry and Hamiltonian information are encoded into a molecule-specific condition $\bm{c}_m$, which modulates both the autoregressive amplitude network and the phase network.
    Given an occupation sequence $\bm{v}$, the amplitude network parameterizes the Born probability $p_{\theta,m}(\bm{v})$, while the phase network predicts $\phi_{\theta,m}(\bm{v})$; together they define the complex wave-function coefficient
    $\Psi_{\theta,m}(\bm{v})=\sqrt{p_{\theta,m}(\bm{v})}\exp[i\phi_{\theta,m}(\bm{v})]$.
    \textbf{b}, FiLM conditioning in the amplitude and phase pathways.
    The molecular condition $\bm{c}_m$ modulates the hidden representations after each Transformer block in the amplitude network and throughout the MLP-based phase network.
    \textbf{c}, Geometry encoder.
    Pairwise nuclear distances are expanded using Gaussian radial basis functions, transformed by an MLP, and aggregated into a fixed-width geometry representation $\bm{z}_{R,m}$.
    \textbf{d}, Hamiltonian encoder.
    The real and imaginary parts of the Hamiltonian coefficients are projected into coefficient tokens, which are pooled by multi-head attention using a learned query $\bm{q}_H$ to produce the fixed-width Hamiltonian representation $\bm{z}_{H,m}$.
    The geometry and Hamiltonian representations are subsequently fused to form $\bm{c}_m$.
    }
    \label{fig:nnqs-model}
\end{figure}

The ansatz in Eqs.~\eqref{eq:autoregressive-born-probability}
and~\eqref{eq:autoregressive-wavefunction} is sufficient for a single fixed Hamiltonian, but an occupation sequence alone does not identify the molecular problem in a shared multi-molecule model.
In particular, the same occupation sequence \(\bm{v}\) can correspond to different wave-function amplitudes and phases under different molecular geometries and Hamiltonians.
We therefore augment the autoregressive ansatz with a molecule-specific condition \(\bm{c}_m\), while sharing the neural-network parameters across all molecular systems.
As illustrated in Fig.~\ref{fig:nnqs-model}a,b, within the physics-conditioned autoregressive NNQS, a geometry encoder and a Hamiltonian encoder construct the molecule-specific condition from molecular geometry and Hamiltonian information, respectively. 
This condition is injected into both the autoregressive amplitude network and the phase network through feature-wise linear modulation (FiLM)~\cite{perez2018film}.

The molecule-specific condition $\bm{c}_m$ is constructed from two complementary sources of molecular information: molecular geometry and the electronic Hamiltonian. The two sources of molecular information are separately mapped to fixed-width representations, which are then fused to form the molecular condition.
The geometry encoder, illustrated in Fig.~\ref{fig:nnqs-model}c, represents molecule \(m\) using the pairwise distances between distinct nuclei:
\begin{equation}
r_{ab}=\lVert\bm{R}_{a}-\bm{R}_{b}\rVert_2, \qquad a<b,
\end{equation}
where \(\bm{R}_{a}\in\mathbb{R}^{3}\) denotes the Cartesian position of nucleus \(a\).
The distances are expanded in 32 Gaussian radial basis functions with centers spanning 0--5~\AA. 
A two-layer multilayer perceptron transforms these features, which are then aggregated into a fixed-width geometry vector \(\bm{z}_{R,m}\). 
Because this vector depends only on pairwise distances and symmetric aggregation, it is unchanged by molecular translation, rotation, or the ordering of the nuclei.

The Hamiltonian encoder, illustrated in Fig.~\ref{fig:nnqs-model}d, encodes the electronic Hamiltonian coefficient \(h_{m,k}\) used in the VMC calculation.
The real and imaginary parts of each Hamiltonian coefficient are encoded as:
\begin{equation}
\bm{e}_{m,k}
=W_H[\operatorname{Re}(h_{m,k});\operatorname{Im}(h_{m,k})]+\bm{b}_H,
\end{equation}
where \(W_H\) and \(\bm{b}_H\) are the learned projection matrix and bias vector, respectively.
Given the resulting variable-length sequence of coefficient tokens
$\bm{E}_m=\{\bm{e}_{m,k}\}_k$, we employ a multi-head attention
(MHA)~\cite{Vaswani2017Attention} pooling mechanism.
Specifically, a learned query $\bm{q}_H$ attends to the coefficient
tokens, which serve as both keys and values:
\begin{equation}
\bm{z}_{H,m}
=
\operatorname{MHA}
\left(
Q=\bm{q}_H,\,
K=\bm{E}_m,\,
V=\bm{E}_m
\right).
\end{equation}
This attention pooling compresses the variable-length set of Hamiltonian
coefficient tokens into a fixed-size representation $\bm{z}_{H,m}$.

The geometry and Hamiltonian representations are then concatenated and fused through a two-layer MLP to obtain the molecule-specific condition:
\begin{equation}
\bm{c}_m=f_{\mathrm{phys}}
\!\left([\bm{z}_{R,m};\bm{z}_{H,m}]\right).
\label{eq:physical-condition}
\end{equation}

Rather than appending $\bm{c}_m$ as an additional occupation token, we inject the molecular condition directly into the hidden representations through FiLM:
\begin{equation}
\widetilde{\bm{h}}^{(\ell)}
=
\left[\bm{1}+\boldsymbol{\gamma}_{\ell}(\bm{c}_m)\right]
\odot\bm{h}^{(\ell)}
+\boldsymbol{\beta}_{\ell}(\bm{c}_m),
\label{eq:physical-film}
\end{equation}
where $\boldsymbol{\gamma}_{\ell}$ and $\boldsymbol{\beta}_{\ell}$
denote learnable projections of $\bm{c}_m$, and $\odot$ denotes element-wise multiplication. 
The FiLM conditioning mechanism is used in both the amplitude and phase pathways to modulate their hidden representations.

\subsection{Molecular benchmarks and model training}
\label{sec:methods-multimolecule-transfer}
We construct two molecular benchmarks of varying difficulty using 12 molecules in the STO-3G basis set. 
The molecules are ranked according to the size of their spin-resolved full configuration interaction (FCI) determinant spaces: the six smallest constitute the Easy benchmark, while the six largest form the Hard benchmark (Table~\ref{tab:molecular-benchmarks}; Supplementary Section~\ref{sec:supp-molecular-systems}).
Thus, the two benchmarks contain the same number of molecular Hamiltonians but differ in the scale of their determinant spaces.

\begin{table}[htbp]
\centering
\renewcommand{\arraystretch}{1.25}
\caption{\textbf{Molecular benchmarks used for pretraining.}}
\label{tab:molecular-benchmarks}
\begin{tabular}{@{}ll@{}} 
\toprule
\textbf{Benchmark} & \textbf{Molecules} \\
\midrule
Easy      & H$_2$O, O$_2$, NH$_3$, CH$_4$, NOH, C$_2$ \\
Hard      & H$_2$O$_2$, HCN, H$_2$CO, N$_2$H$_2$, NOH$_3$, C$_2$H$_2$ \\
\bottomrule
\end{tabular}
\end{table}

For either benchmark, let \(\mathcal D\) denote its set of molecules. 
The molecule-specific condition allows a single set of network parameters to be shared across all \(m\in\mathcal D\). 
The VMC energy \(E_m(\theta)\) is estimated for each molecule, and the model is optimized using their uniformly weighted average:
\begin{equation}
J_{\mathcal{D}}(\theta)
=
\frac{1}{|\mathcal{D}|}
\sum_{m\in\mathcal{D}}E_m(\theta).
\label{eq:multimolecule-objective}
\end{equation}
Thus, every molecule contributes equally to the pretraining objective, regardless of the size of its determinant space.
Notably, models for the Easy and Hard benchmarks are independently pretrained from scratch, with no weight transfer between them.

\subsection{Evaluated configurations and compute estimation}
\label{sec:methods-model-compute}

We evaluate a grid defined by trainable parameter count \(P\) and cumulative optimization steps \(S\), comprising eight model sizes with approximately 0.1M, 0.2M, 0.4M, 0.9M, 1.5M, 3.5M, 5.8M and 13.8M parameters, and five optimization milestones at 10K, 20K, 30K, 40K and 50K steps. 
Following a compound-scaling strategy~\cite{Zhai2022Scaling}, the model sizes are constructed by varying network width and depth while retaining the same overall ansatz design (Table~\ref{tab:model-ladder}; Supplementary Section~\ref{sec:supp-architecture}). 

\begin{table}[htbp]
\centering
\caption{\textbf{Model family used in the scaling experiments.} Here, amplitude width denotes the hidden dimension of the amplitude Transformer, and phase width denotes the hidden dimension of each layer in the phase MLP. Full architectural and conditioning details are provided in Supplementary Section~\ref{sec:supp-architecture}.}
\label{tab:model-ladder}
\begin{tabular}{ccccr}
\toprule
Scale & Amplitude width & Transformer blocks & Phase width & Parameters \\
\midrule
1 & 32  & 4 & 32  & 97,763 \\
2 & 32  & 8 & 64  & 226,611 \\
3 & 64  & 4 & 64  & 375,747 \\
4 & 64  & 8 & 128 & 880,227 \\
5 & 128 & 4 & 128 & 1,472,387 \\
6 & 128 & 8 & 256 & 3,468,483 \\
7 & 256 & 4 & 256 & 5,828,355 \\
8 & 256 & 8 & 512 & 13,769,091 \\
\bottomrule
\end{tabular}
\end{table}

Estimating computational cost is essential for translating scaling behavior into practical resource-allocation decisions.
Although model size ($P$) and optimization step ($S$) define a training configuration, they serve only as indirect descriptors and fail to capture the true cumulative computational demand. 
Therefore, we mapped each $(P,S)$ pair to a unified computational metric. 
A basic approximation is the product of parameter count and update count, \(P\times S\). 
However, this approximation treats every update as having the same sampling workload.
In our VMC implementation, this workload was not constant: MCTS sampling retained different numbers of distinct configurations at different steps after repeated samples were combined, and configuration lengths differed across molecules.
To address this, we utilized logged configuration counts and sequence lengths to establish a cumulative, token-weighted sampling workload.

For molecule \(m\), let \(U_m(P,S)\) denote the cumulative count of sampled configurations, obtained by summing the counts of distinct configurations retained at each update while training a model of size \(P\) through milestone \(S\).
Let \(T_m\) denote the length of the input occupation sequence for
molecule \(m\). 
The product \(U_m(P,S)T_m\) hence estimates the cumulative
sampled-token workload for molecule \(m\) through milestone \(S\). The total cumulative workload across the benchmark set \(\mathcal{D}\) is:
\begin{equation}
\Omega(P,S)
=\sum_{m\in\mathcal{D}}U_m(P,S)T_m.
\label{eq:token-weighted-terminal-workload}
\end{equation}

Finally, we estimated cumulative compute by combining the cost of processing the sampled configurations, \(\Omega(P,S)\), with the cost of updating the model parameters during training, \(P \times S\):
\begin{equation}
C(P,S)=P\times(\lambda_{1}\Omega(P,S)+\lambda_{2}S).
\label{eq:known-compute-proxy}
\end{equation}
Here, we adopt \(\lambda_1=8\) and \(\lambda_2=12\) in this work (see Supplementary Section~\ref{sec:supp-compute-accounting} for details).
Notably, this hardware-independent proxy is derived from logged workload statistics rather than hardware counters and does not represent a complete end-to-end FLOP count.

\subsection{Error metrics and compute-efficient configuration selection}
\label{sec:methods-scaling-analysis}

For molecule \(m\), the error at milestone \(S\) is defined as the mean absolute deviation from the reference energy over the final \(W=1,000\) updates:
\begin{equation}
\epsilon_m(P,S)
=\frac{1}{W}\sum_{t=1}^{W}
\left|\widehat E_{m,t}(P,S)-E_m^{\mathrm{ref}}\right|,
\label{eq:molecular-error-metric}
\end{equation}
where $t$ denotes the update index within the evaluation window ending at $S$. 
The overall benchmark error is then calculated by taking an unweighted average across all constituent molecules:
\begin{equation}
\mathcal{L}_{\mathcal{D}}(P,S)
=\frac{1}{|\mathcal{D}|}\sum_{m\in\mathcal{D}}\epsilon_m(P,S).
\label{eq:joint-error-metric}
\end{equation}

To identify compute-efficient choices, we compare all evaluated \((P,S)\) pairs by their error and cumulative compute. A pair is Pareto efficient if no other evaluated pair achieves both lower error and lower compute. For a compute budget \(C_{\max}\), we select:
\begin{equation}
(P^*,S^*)
=
\underset{C(P,S)~\leq~C_{\max}}{\operatorname{arg\,min}}\;
\mathcal{L}_{\mathcal{D}}(P,S).
\label{eq:methods-compute-optimal-decision}
\end{equation}
Thus, the decision rule selects the lowest-error configuration within the measured grid that satisfies the compute budget.

\subsection{Implementation details}
\label{sec:methods-implementation}

The models are implemented in PyTorch and optimized in FP32 precision using the AdamW optimizer. For the pretraining and fine-tuning runs, we use a learning rate of $3 \times 10^{-4}$, $\beta = (0.9, 0.99)$, $\epsilon = 10^{-9}$, and zero weight decay. 
Local energies are evaluated using a compiled C++/CUDA backend. Pauli strings sharing the same occupation-flip pattern are grouped together, while electronic configurations and flip patterns are encoded as 64-bit blocks to facilitate parallel lookup and evaluation. Further sampling, optimization and reproducibility details are provided in Supplementary Section~\ref{sec:supp-vmc-optimization}.

\bibliographystyle{unsrt}
\bibliography{references}

\vspace{36pt}

\section*{Author contributions} 
W.W. and H.S. conceived the research idea and designed the study. C.Y., H.K., J.W., and L.F. developed the neural network, conducted the experiments, and prepared the initial draft. W.W., H.S. and J.Y. provided critical revisions and refined the technical presentation of the manuscript. W.W. and H.S. supervised the project. All authors reviewed and approved the final manuscript.

\section*{Corresponding authors} 
Honghui Shang (\url{shanghui.ustc@gmail.com}) and Wenguan Wang (\url{wenguanwang@zju.edu.cn}).

\section*{Competing interests}
The authors declare no competing interests.

\clearpage
\pdfbookmark[1]{Supplementary Information}{supplementary-information}
\section*{Supplementary Information}
\setcounter{subsection}{0}
\renewcommand{\thesubsection}{S\arabic{subsection}}
\renewcommand{\theHsubsection}{suppmethod.\arabic{subsection}}
\setcounter{figure}{0}
\renewcommand{\thefigure}{S\arabic{figure}}
\renewcommand{\theHfigure}{supp.\arabic{figure}}
\renewcommand{\figurename}{Supplementary Figure}
\setcounter{table}{0}
\renewcommand{\thetable}{S\arabic{table}}
\renewcommand{\theHtable}{supp.\arabic{table}}
\renewcommand{\tablename}{Supplementary Table}
\setcounter{equation}{0}
\renewcommand{\theequation}{S\arabic{equation}}
\renewcommand{\theHequation}{supp.\arabic{equation}}

\subsection{Molecular Hamiltonians and benchmark construction}
\label{sec:supp-molecular-systems}

Molecular Hamiltonians are generated within the Born--Oppenheimer approximation using PySCF 2.1.1~\cite{Sun2018PySCF} and OpenFermion 1.5.1~\cite{McClean2020OpenFermion}. All source molecules are neutral and described in the STO-3G basis~\cite{Hehre1969STO}. All electrons are retained, with no frozen-core approximation or active-space truncation. Except for O$_2$, the molecules are treated in closed-shell sectors with \(N_\alpha=N_\beta\). O$_2$ is treated in the triplet \(M_S=1\) sector, corresponding to \((N_\alpha,N_\beta)=(9,7)\). Canonical Hartree--Fock orbitals are retained in PySCF order, and the resulting second-quantized Hamiltonians are mapped to qubits by the Jordan--Wigner transformation. Reference energies are computed in the same orbital spaces. Reference values and their sources are listed in Supplementary Table~\ref{tab:hamiltonians}.

The FCI determinant-space size is used to define the reported benchmarks. For a molecule with \(M\) spatial orbitals and spin-resolved electron counts \((N_\alpha,N_\beta)\), the determinant count is:
\begin{equation}
    N_{\mathrm{det}}=\binom{M}{N_\alpha}\binom{M}{N_\beta}.
\end{equation}
The molecules used in the reported analyses are ordered by \(N_{\mathrm{det}}\) and assigned to two contiguous groups of six molecules, denoted Easy and Hard.

The source Hamiltonians are generated once and reused across all model sizes and seeds. 
The held-out N$_2$ target uses a bond length of 1.0784~\AA, with the two nuclei placed symmetrically about the origin, and is absent from all source-pretraining inputs and model-selection runs.

\begin{table}[htbp]
\centering
\caption{\textbf{Molecular Hamiltonians and reference energies.}
All energies are in Hartree and the reference energies are obtained from FCI.
$N_{\mathrm{det}}$ denotes the size of the spin-resolved FCI determinant space
in the STO-3G basis. Cartesian coordinates are provided in Supplementary Data~1.}
\label{tab:hamiltonians}
\resizebox{0.85\textwidth}{!}{
\begin{tabular}{llcrrrr}
\toprule
Benchmark & Molecule & $(N_\alpha,N_\beta)$ & Orbitals
& Qubits & $N_{\mathrm{det}}$ & $E_{\mathrm{ref}}$ (Ha) \\
\midrule
\multirow{6}{*}{\textbf{Easy}}
& H$_2$O       & $(5,5)$ & 7  & 14 & 441     & -75.0063982800 \\
& O$_2$        & $(9,7)$ & 10 & 20 & 1,200   & -147.7216652678 \\
& NH$_3$       & $(5,5)$ & 8  & 16 & 3,136   & -55.5151425685 \\
& CH$_4$       & $(5,5)$ & 9  & 18 & 15,876  & -39.8045269353 \\
& NOH          & $(8,8)$ & 11 & 22 & 27,225  & -128.2169318028 \\
& C$_2$        & $(6,6)$ & 10 & 20 & 44,100  & -74.6907819191 \\
\midrule
\multirow{6}{*}{\textbf{Hard}}
& H$_2$O$_2$   & $(9,9)$ & 12 & 24 & 48,400  & -148.8551873571 \\
& HCN          & $(7,7)$ & 11 & 22 & 108,900 & -91.8335485935 \\
& H$_2$CO      & $(8,8)$ & 12 & 24 & 245,025 & -112.3877360240 \\
& N$_2$H$_2$   & $(8,8)$ & 12 & 24 & 245,025 & -108.6987774392 \\
& NOH$_3$      & $(9,9)$ & 13 & 26 & 511,225 & -129.3602760342 \\
& C$_2$H$_2$   & $(7,7)$ & 12 & 24 & 627,264 & -76.0247473113 \\
\midrule
\textbf{Transfer}
& N$_2$        & $(7,7)$ & 10 & 20 & 14,400  & -107.6402331522 \\
\bottomrule
\end{tabular}
}
\end{table}

\subsection{Conditional network architecture}
\label{sec:supp-architecture}

The autoregressive amplitude–phase factorization and the integration of the molecular condition are detailed in Methods Sections \ref{sec:methods-autoregressive-ansatz} and \ref{sec:methods-physical-conditioning}. 
In this section, we describe the implementation details common to all models evaluated in this study.

The amplitude network employs a decoder-only causal Transformer architecture with pre-layer normalization. 
Each Transformer block consists of a four-head self-attention mechanism followed by a GELU feed-forward network of width $4d$, utilizing residual connections around both sublayers without dropout.
The phase network is implemented as a multilayer perceptron (MLP) featuring four hidden layers with ReLU activations, which processes the entire occupation sequence. 
To strictly preserve the target $N_\alpha$ and $N_\beta$ sectors during autoregressive sampling, invalid occupation tokens are explicitly masked prior to normalization.

The amplitude and phase pathways share a common geometry encoder and Hamiltonian encoder, whose parameters are jointly used by both pathways.
Within the amplitude network, FiLM is applied after each Transformer block. 
In the phase network, FiLM is applied after each hidden linear layer of the MLP.

\subsection{VMC sampling and optimization}
\label{sec:supp-vmc-optimization}

Here, we provide additional details of the gradient estimator and adaptive sampling procedure.
For each molecule, the energy gradient is estimated with the standard VMC estimator:
\begin{equation}
\nabla_\theta E_m
=
2\operatorname{Re}\,
\mathbb{E}_{\bm{v}\sim p_{\theta,m}}
[
(E_{\mathrm{loc},m}(\bm{v})-E_m)
\nabla_\theta\log\Psi_{\theta,m}^*(\bm{v})
],
\end{equation}
where \(\theta\) denotes the variational parameters, \(\bm{v}\) is an occupation configuration sampled from the Born distribution \(p_{\theta,m}(\bm{v})=|\Psi_{\theta,m}(\bm{v})|^2\), and \(E_m\) is the corresponding variational energy for molecule \(m\). 
The local energy \(E_{\mathrm{loc},m}(\bm{v})\) is defined in Eq.~\eqref{eq:local-energy}, the superscript \(^*\) denotes complex conjugation, and \(\operatorname{Re}\) takes the real part.
For multi-molecule pretraining, the per-molecule gradients are combined with equal weight, consistent with the objective \(J_{\mathcal D}\) defined in Eq.~\eqref{eq:multimolecule-objective}.

Both the pretraining and fine-tuning stages employ the optimizer, learning rate, and numerical precision detailed in Table~\ref{tab:optimization}, utilizing a constant learning-rate schedule. Furthermore, these optimization configurations are maintained consistently across the entire evaluated model grid.

\begin{table}[htbp]
\centering
\caption{\textbf{NNQS optimization settings for source pretraining and N$_2$ fine-tuning.}
Both stages use AdamW with a constant learning-rate schedule and FP32 precision. 
The two stages differ primarily in the target number of unique sampled configurations and the total number of optimization steps.}
\label{tab:optimization}
\begin{tabular}{p{0.25\textwidth}p{0.3\textwidth}p{0.2\textwidth}}
\toprule
Setting & Source pretraining & N$_2$ fine-tuning \\
\midrule
Optimizer & AdamW & AdamW \\
Learning rate & \(3\times10^{-4}\) & \(3\times10^{-4}\) \\
\(\beta_1,\beta_2\) & \((0.9,0.99)\) & \((0.9,0.99)\) \\
Epsilon & \(10^{-9}\) & \(10^{-9}\) \\
Schedule & Constant & Constant \\
Precision & fp32 & fp32 \\
Sampling target & \(10^2\)--\(10^4\) unique & \(10^4\) unique \\
Optimization steps & 50,000 & 10,000 \\
\bottomrule
\end{tabular}
\end{table}

When evaluating the energy and gradient estimators, repeated terminal configurations are stored and evaluated only once, with their integer multiplicities retained as Monte Carlo weights. This avoids redundant computation while preserving the same energy and gradient estimates as explicitly retaining every sampled copy.

During source pretraining, the nominal number of sampling draws is adapted according to the number of unique terminal configurations obtained in the preceding sampling call. If this number falls below the target range, the draw count is multiplied by 1.3; if it exceeds the target range, the draw count is divided by 1.3. The adjusted value is then used to initialize the next sampling call. For N$_2$ fine-tuning, the nominal draw count is adjusted analogously using a factor of 1.2.

\subsection{Power-law formulation and evaluation}
\label{sec:supp-joint-scaling}

In this section, we formalize the separable and interaction power laws and evaluate them against a more flexible quadratic log-response surface. 
The Easy and Hard benchmarks are fitted independently. 
Each fit uses the complete grid of eight model sizes and five optimization milestones, giving \(n=40\) observations per benchmark.
For observation \(i\), we define the dimensionless log-error and the two normalized log-resources as:
\begin{equation}
y_i
=\log(\frac{\mathcal L_{\mathcal D}(P_i,S_i)}{1\,\mathrm{mHa}}),
\qquad
x_i=\log(\frac{P_i}{P_{\mathrm{ref}}}),
\qquad
z_i=\log(\frac{S_i}{S_{\mathrm{ref}}}),
\label{eq:supp-log-scaling-variables}
\end{equation}
where \(P_{\mathrm{ref}}=10^6\) parameters and \(S_{\mathrm{ref}}=10^4\) optimization steps. We compare three candidate models:
\begin{align}
\text{separable:}\qquad
\widehat y_i
&=c_0-a x_i-b z_i,
\nonumber\\
\text{interaction:}\qquad
\widehat y_i
&=c_0-a x_i-b z_i-\eta x_i z_i,
\nonumber\\
\text{quadratic:}\qquad
\widehat y_i
&=c_0+c_1x_i+c_2z_i+c_3x_i^2+c_4x_i z_i+c_5z_i^2.
\label{eq:supp-scaling-candidate-models}
\end{align}
The separable model assumes additive effects of model size and optimization steps in log space. The interaction model adds the single cross term \(x_i z_i\), allowing the effect of one resource to depend on the other. The quadratic model is more flexible because it additionally allows curvature along both resource axes; no sign constraints are imposed on its coefficients.

Having specified the three candidate response surfaces, we estimate their coefficients separately for each benchmark by ordinary least squares in log space.
For a candidate model \(M\), the fitted coefficients minimize the residual sum of squares (RSS):
\begin{equation}
\mathrm{RSS}_{M}
=\sum_{i=1}^{n}\left(y_i-\widehat y_{i,M}\right)^2.
\label{eq:supp-scaling-rss}
\end{equation}
Here, \(y_i-\widehat y_{i,M}\) is the log-space residual for observation \(i\), namely the difference between the observed and predicted log-errors. 
Equivalently, it is the logarithm of the ratio between the observed and predicted energy errors. 
A smaller RSS therefore indicates closer agreement with the measured grid.

To assess and compare the adequacy of these three candidate response surfaces, we use three complementary criteria: log-space \(R^2\) for in-sample agreement, AICc for the trade-off between fit quality and model complexity, and grouped cross-validation for held-out predictive performance.

The in-sample goodness-of-fit, measured relative to a mean-only baseline, is captured by the log-space coefficient of determination:
\begin{equation}
R^2_{\log,M}
=1-
\frac{\mathrm{RSS}_{M}}
{\sum_{i=1}^{n}(y_i-\overline y)^2},
\qquad
\overline y=\frac{1}{n}\sum_{i=1}^{n}y_i.
\label{eq:supp-scaling-r2}
\end{equation}
A larger \(R^2_{\log}\) indicates closer in-sample agreement. 
However, adding coefficients cannot reduce the best attainable \(R^2_{\log}\), so this statistic alone may favor an unnecessarily flexible model. 

We therefore next use the small-sample corrected Akaike information criterion (AICc) to balance fit quality against model complexity. Under the Gaussian log-residual model used with ordinary least squares, and omitting an additive constant common to all candidates:
\begin{equation}
\mathrm{AICc}_{M}
=n\log (\frac{\mathrm{RSS}_{M}}{n})
+2k_M
+\frac{2k_M(k_M+1)}{n-k_M-1},
\label{eq:supp-scaling-aicc}
\end{equation}
where $k_M$ denotes the total number of estimated parameters, comprising both the regression coefficients and the fitted residual variance. 
Accordingly, $k_M = 4$, $5$, and $7$ for the separable, interaction, and quadratic models, respectively. 
A lower AICc value indicates a better trade-off between in-sample fit and model complexity. 
Note that AICc differences are only meaningful when comparing models fitted to the same benchmark.

Although AICc penalizes additional parameters, it is still calculated from the same observations used to fit each model. 
As a complementary test of predictive robustness, we therefore use grouped cross-validation along the two resource axes~\cite{Stone1974CrossValidation,Roberts2017GroupedCrossValidation}. 
CV-P evaluates prediction at an excluded model size, whereas CV-S evaluates prediction at an excluded optimization milestone. 
Both are reported as root-mean-square errors in natural-log units, with lower values indicating better held-out prediction.
Specifically, for candidate model \(M\),
\begin{equation}
\begin{aligned}
\mathrm{CV\text{-}P}_{M}
&=[\frac{1}{n}\sum_{i=1}^{n}
(y_i-\widehat y_{i,M}^{(-P_i)})^2]^{1/2},\\
\mathrm{CV\text{-}S}_{M}
&=[\frac{1}{n}\sum_{i=1}^{n}
(y_i-\widehat y_{i,M}^{(-S_i)})^2]^{1/2}.
\end{aligned}
\label{eq:supp-scaling-cv}
\end{equation}
Here, \(\widehat y_{i,M}^{(-P_i)}\) is the prediction for observation \(i\) from a model fitted after excluding all observations at model size \(P_i\); \(\widehat y_{i,M}^{(-S_i)}\) is defined analogously by excluding optimization milestone \(S_i\).

As shown in Table~\ref{tab:joint-model-selection}, the interaction power law is preferred by AICc for both benchmarks. 
Relative to the separable model, it also improves held-out prediction along both the model-size and optimization-step axes. 
The more flexible quadratic response surface achieves slightly higher in-sample \(R^2_{\log}\), but is disfavored by AICc and shows substantially larger CV-P errors. 
The fitted interaction is stronger for Easy than for Hard, with \(\eta=0.3777\) and \(0.1325\), respectively (Supplementary Table~\ref{tab:joint-fit-coefficients}). 
Because multiple milestones are obtained from the same training trajectory, these comparisons are interpreted descriptively.

\begin{table}[htbp]
\centering
\caption{\textbf{Comparison of candidate joint scaling models for the Easy and Hard benchmarks.}
The separable power law, interaction power law, and quadratic log-response surface are evaluated using the log-space coefficient of determination \(R^2_{\log}\), the small-sample corrected Akaike information criterion (AICc), and grouped cross-validation.
Higher \(R^2_{\log}\) indicates better in-sample agreement, whereas lower AICc indicates a better trade-off between fit quality and model complexity.
CV-P and CV-S report root-mean-square prediction errors in natural-log units when each model size or optimization milestone, respectively, is held out in turn; lower values indicate better held-out predictive performance.}
\label{tab:joint-model-selection}
\begin{tabular}{llrrrr}
\toprule
Benchmark & Model & \(R^2_{\log}\) & AICc & CV-P & CV-S \\
\midrule
Easy & Separable power law & 0.737 & -35.09 & 0.709 & 0.636 \\
Easy & Interaction power law & 0.829 & -49.70 & 0.629 & 0.478 \\
Easy & Quadratic log surface & 0.837 & -45.82 & 0.762 & 0.484 \\
Hard & Separable power law & 0.912 & -96.14 & 0.351 & 0.287 \\
Hard & Interaction power law & 0.930 & -102.36 & 0.325 & 0.250 \\
Hard & Quadratic log surface & 0.935 & -99.82 & 0.485 & 0.245 \\
\bottomrule
\end{tabular}
\end{table}

Finally, the one-axis exponents shown in Fig.~\ref{fig:conditional-exponents} are obtained from separate regressions at fixed values of the complementary resource:
\begin{equation}
y=c_j-\widetilde\alpha_P(S_j)x
\quad\text{at fixed }S_j,
\qquad
y=d_k-\widetilde\alpha_S(P_k)z
\quad\text{at fixed }P_k.
\label{eq:supp-slice-exponents}
\end{equation}
These slice-wise exponents are descriptive estimates obtained independently of the joint interaction fit and should not be interpreted as coefficients of the interaction model.

\begin{table}[htbp]
\centering
\caption{\textbf{Fitted coefficients of the interaction scaling model for the Easy and Hard benchmarks.}
Coefficients correspond to Eq.~\eqref{eq:joint-interaction-scaling}, with reference scales \(P_{\mathrm{ref}}=10^6\) parameters and \(S_{\mathrm{ref}}=10^4\) optimization steps. 
\(L_0\) denotes the fitted energy error at the reference scales, \(a\) and \(b\) quantify the model-size and optimization-step dependences, respectively, and \(\eta\) measures their coupling. 
\(R^2_{\log}\) reports the coefficient of determination in log space.}
\label{tab:joint-fit-coefficients}
\begin{tabular}{lrrrrr}
\toprule
Benchmark & \(L_0\) (mHa) & \(a\) & \(b\) & \(\eta\) & \(R^2_{\log}\) \\
\midrule
Easy & 17.56 & 0.0618 & 1.164 & 0.3777 & 0.829 \\
Hard & 13.66 & 0.3367 & 0.7826 & 0.1325 & 0.930 \\
\bottomrule
\end{tabular}
\end{table}

\subsection{Compute-proxy decomposition}
\label{sec:supp-compute-accounting}

In this section, we explain how the compute proxy in Eq.~\eqref{eq:known-compute-proxy} of Methods Section~\ref{sec:methods-model-compute} is constructed and converted to EFLOP-proxy units. 
By combining the cost of processing the sampled configurations, denoted by $\Omega(P,S)$, with the cost of updating $P$ model parameters across $S$ optimization steps, we obtain the total compute:
\begin{equation}
C(P,S)
=
\underbrace{6P\times\Omega(P,S)}_{\text{forward and backward}}
+
\underbrace{2P\times\Omega(P,S)}_{\text{autoregressive sampling}}
\quad+
\underbrace{12P\times S}_{\text{AdamW updates}}.
\label{eq:known-compute}
\end{equation}
The coefficients follow a simple operation-counting convention, \textit{i.e.}, one multiplication and one accumulation.
A forward evaluation is approximated as \(2P\) operations per token, and the backward pass adds approximately \(4P\), giving the coefficient \(6\) for a gradient-bearing evaluation. 
Autoregressive sampling requires a separate forward pass without gradient tracking to generate the conditional token probabilities, contributing the coefficient \(2\). 
The first-moment update, second-moment update, and parameter update in AdamW are approximated together as \(12P\) operations per training step. 

\subsection{Pareto-frontier fitting}
\label{sec:supp-pareto-fitting}

In this section, we explain how the empirical error-compute Pareto frontier is constructed and fitted. 
For each benchmark, we compare the compute and error of all evaluated model-milestone configurations. 
A configuration is excluded if another evaluated configuration uses no more compute and achieves no larger error, with at least one of these quantities being strictly smaller. 
The remaining non-dominated configurations form the empirical Pareto frontier.

To quantify how quickly the frontier error decreases with compute, we fit the Pareto points in log--log space. 
For Pareto point \(i\), we define the log compute and dimensionless log error as:
\begin{equation}
y_i=\log(\frac{\mathcal L_i}{1\,\mathrm{mHa}}),
\qquad
u_i=\log C_i.
\label{eq:supp-pareto-variables}
\end{equation}
where \(C_i\) is the compute proxy and \(\mathcal L_i\) is the measured benchmark error. 
To this end, we compare three candidate curves:
\begin{align}
\text{single power:}\qquad
\widehat y_i
&=a-bu_i,
\nonumber\\
\text{quadratic log:}\qquad
\widehat y_i
&=a+bu_i+du_i^2,
\nonumber\\
\text{broken power:}\qquad
\widehat y_i
&=
\begin{cases}
a-b_{<}(u_i-u_c), & u_i\leq u_c,\\
a-b_{>}(u_i-u_c), & u_i>u_c,
\end{cases}
\label{eq:supp-pareto-candidate-models}
\end{align}
where $\widehat y_i$ is the fitted value of \(y_i\), and $a$, $b$, and $d$ are estimated coefficients. 
The parameter $u_c$ is the change point, with $b_{<}$ and $b_{>}$ denoting the slopes below and above it.

Each candidate curve is fitted to the measured Pareto points by least squares in log space. 
We compare the three curves using AICc, defined in Eq.~\eqref{eq:supp-scaling-aicc}; a lower AICc indicates a better balance between fitting the data and avoiding unnecessary parameters. 
Unlike the analysis in Supplementary Section~\ref{sec:supp-joint-scaling}, which uses the complete \(P\)--\(S\) grid to study the joint effects of model size and optimization step, this analysis uses only the Pareto points and asks how the lowest measured error changes with total compute.
The fitted frontier curves serve solely to summarize the observed trend.
As shown in Table~\ref{tab:pareto-model-selection}, the single-power curve is preferred by AICc for the Easy benchmark, whereas the Hard benchmark is best characterized by the broken-power curve.

\begin{table}[htbp]
\centering
\caption{\textbf{Pareto-frontier model comparison by AICc.} Lower values indicate the preferred model for each benchmark, with the best-performing model highlighted in bold.}
\label{tab:pareto-model-selection}
\begin{tabular}{@{}cc|ccc@{}}
\toprule
Benchmark & Pareto Points & Single Power & Quadratic Log & Broken Power \\
\midrule
Easy & 18 & $\mathbf{-27.03}$ & $-24.12$ & $-21.25$ \\
Hard & 21 & $-42.17$ & $-60.61$ & $\mathbf{-69.74}$ \\
\bottomrule
\end{tabular}
\end{table}

Let $\widetilde C$ denote the estimated computational cost, measured in EFLOP, and let $\widehat{\mathcal L}^{*}(\widetilde C)$ denote the fitted error along the Pareto frontier.
For the Easy benchmark, the single-power model is expressed as:
\begin{equation}
\widehat{\mathcal L}^{*}_{\mathrm{Easy}}(\widetilde C)
=
3.44
(\frac{\widetilde C}{0.01})^{-0.512}
\,\mathrm{mHa}.
\label{eq:supplementary-compute-frontier-easy}
\end{equation}
For the Hard benchmark, the corresponding broken-power model is formulated as:
\begin{equation}
\widehat{\mathcal L}^{*}_{\mathrm{Hard}}(\widetilde C)
=
6.84\,\mathrm{mHa}
\begin{cases}
(\dfrac{\widetilde C}{0.00886})^{-1.062},
& \widetilde C\leq0.00886,\\[3pt]
(\dfrac{\widetilde C}{0.00886})^{-0.367},
& \widetilde C>0.00886.
\end{cases}
\label{eq:supplementary-compute-frontier-hard}
\end{equation}

\subsection{Transfer initialization and evaluation}
\label{sec:supp-n2-transfer}

In this section, we first specify the initialization for fine-tuning. 
We retain all pretrained weights with matching names and shapes, except for the phase network and amplitude output head, which are reinitialized. 
If the sequence lengths differ, the shared prefix of the position embeddings is copied.
All optimization and training states are reset prior to the target runs.

We then provide a more detailed analysis of the experiments presented in Section~\ref{results:transfer_N2}, further assessing whether the favorable model-size trend observed during source pretraining persists after fine-tuning on the held-out target.
For each initialization, we fit:
\begin{equation}
\widehat{\mathcal L}(P)
=
A(\frac{P}{10^6})^{-\alpha},
\label{eq:supp-transfer-scaling}
\end{equation}
where \(P\) is the number of trainable parameters, \(A\) is the fitted error at one million parameters and \(\alpha\) describes the model-size trend. A positive \(\alpha\) indicates decreasing error with model size, whereas a negative \(\alpha\) indicates the opposite fitted direction.

The Easy- and Hard-pretrained runs yield \(\alpha=0.259\) and \(0.447\), respectively, showing decreasing target error with model size; the larger Hard-pretrained exponent indicates a steeper decrease. 
The scratch fit yields \(\alpha=-0.272\), but its low fit quality (\(R^2_{\log}=0.294\)) does not support a stable reverse-scaling trend. 
Instead, the scratch results show no clear decrease in error with model size under the fixed fine-tuning budget. 
These fits are based on independently selected runs drawn from equal candidate pools.

\begin{table}[htbp]
\centering
\caption{\textbf{Descriptive N$_2$ parameter-scaling fits.} Parameters correspond to Eq.~\eqref{eq:supp-transfer-scaling}, with \(A\) reported in mHa.}
\label{tab:n2-parameter-fits}
\begin{tabular}{lrrr}
\toprule
Initialization & \(A\) (mHa) & \(\alpha\) & \(R^2_{\log}\) \\
\midrule
Easy pretrained & 0.726 & 0.259 & 0.607 \\
Hard pretrained & 0.457 & 0.447 & 0.868 \\
Scratch & 8.23 & -0.272 & 0.294 \\
\bottomrule
\end{tabular}
\end{table}

\subsection{Conditional parameter scaling across optimization milestones}
\label{sec:supp-conditional-scaling}

\begin{figure}[!ht]
  \centering
  \includegraphics[width=0.8\textwidth]{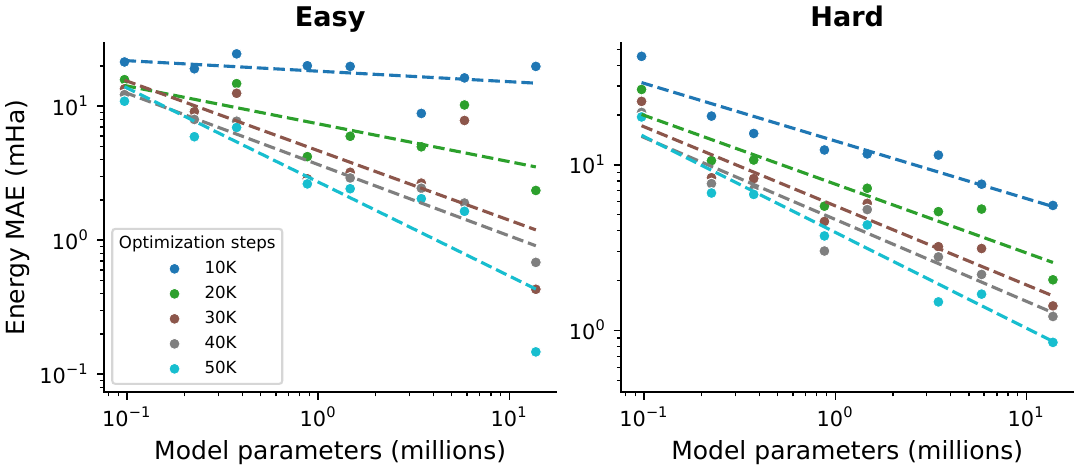}
  \caption{\textbf{Benchmark-level parameter scaling across optimization milestones.} The equally weighted mean absolute energy error (MAE, mHa) across molecules is plotted against the number of trainable model parameters for the Easy (left) and Hard (right) benchmarks. Colors denote cumulative optimization milestones from 10K to 50K optimization steps. Points show the evaluated model sizes, and dashed lines show separate descriptive power-law fits at each fixed milestone. Both axes are logarithmic, and the fitted lines summarize only the measured parameter range.}
  \label{fig:supp-benchmark-parameter-scaling}
\end{figure}

Supplementary Fig.~\ref{fig:supp-benchmark-parameter-scaling} summarizes the benchmark-level dependence of MAE on model size at each optimization milestone. 
The fitted decrease becomes steeper with continued optimization on both benchmarks, whereas Easy retains greater variation among individual model scales.

\begin{figure}[!ht]
  \centering
  \includegraphics[width=0.82\textwidth]{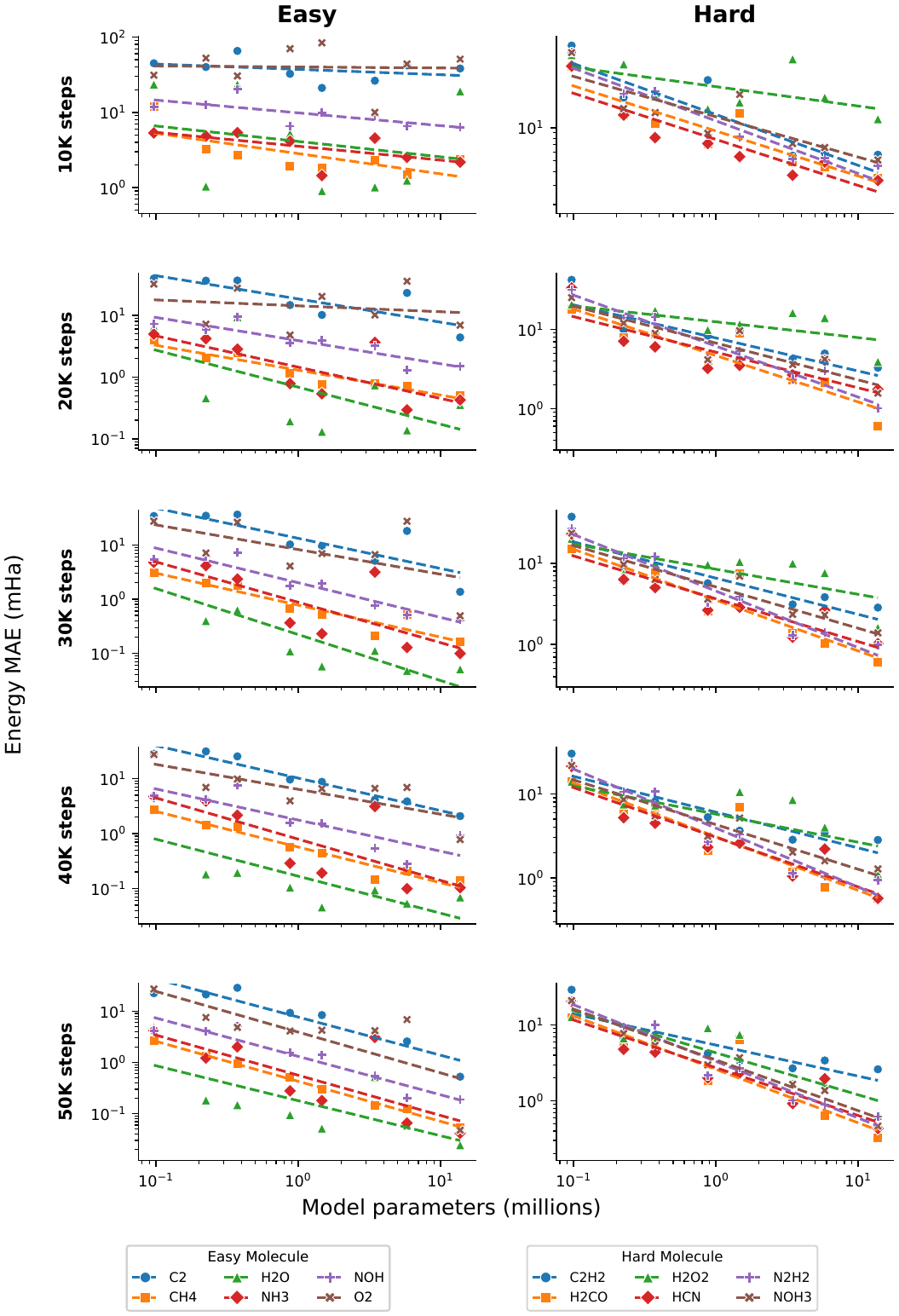}
  \caption{\textbf{Molecule-resolved parameter scaling across pretraining stages.} Molecule-level MAE is plotted against the number of trainable model parameters for the six molecules in the Easy (left) and Hard (right) benchmarks at different pretraining stages. From top to bottom, the rows correspond to 10K, 20K, 30K, 40K, and 50K optimization steps. Symbols and colors identify individual molecules. Dashed lines show separate descriptive power-law fits for each molecule at each pretraining stage over the evaluated model sizes.}
  \label{fig:supp-molecule}
\end{figure}

At 10K and 20K steps, Hard already shows decreasing fitted trends across its constituent molecules (Supplementary Fig.~\ref{fig:supp-molecule}). 
Easy is more heterogeneous, with weak or non-monotonic molecule-level responses superimposed on the benchmark-level trend.

By 30K steps, decreasing fitted trends are visible for most molecules in both benchmarks, although isolated departures from monotonicity remain. 
The aggregate capacity trend therefore does not require identical checkpoint ordering for every molecule.

The 40K- and 50K-update profiles show the clearest molecule-level capacity trends within the evaluated range. 
Residual scatter remains molecule dependent, so these lines are descriptive finite-range fits rather than universal scaling laws.

\clearpage
\subsection{Ablation of the physics-conditioned autoregressive NNQS}

\begin{table*}[h]
    \centering
    \caption{
        Ablation of conditioning in the physics-conditioned autoregressive NNQS on the Easy benchmark. Values are mean absolute energy errors (mHa) over six
        molecules, averaged over the 1,000-update window ending at each optimization milestone.
        The last column reports the absolute error increase and degradation
        factor after removing conditioning at 50K steps.
        The ablation uses the four-block models with amplitude widths of 32, 64, 128, and 256.
    }
    \label{tab:physics_encoding_ablation}

    \small
    \setlength{\tabcolsep}{5.5pt}
    \renewcommand{\arraystretch}{1.08}

    \begin{tabular*}{\textwidth}{
        @{\extracolsep{\fill}}
        c l
        r r r r r
        c
    }
        \toprule
        \multirow{2}{*}{Width}
        & \multirow{2}{*}{Encoding}
        & \multicolumn{5}{c}{MAE (mHa) $\downarrow$}
        & \multirow{2}{*}{50K degradation} \\
        \cmidrule(lr){3-7}
        & & 10K & 20K & 30K & 40K & 50K & \\

        \midrule

        \multirow{2}{*}{32}
        & with
        & \textbf{21.406}
        & \textbf{15.796}
        & \textbf{13.488}
        & \textbf{12.211}
        & \textbf{10.908}
        & -- \\
        & without
        & 75.815 & 74.940 & 74.948 & 74.713 & 74.498
        & +63.590 ($6.83\times$) \\

        \addlinespace[2pt]

        \multirow{2}{*}{64}
        & with
        & \textbf{24.606}
        & \textbf{14.757}
        & \textbf{12.499}
        & \textbf{7.735}
        & \textbf{6.943}
        & -- \\
        & without
        & 75.572 & 74.420 & 74.050 & 73.796 & 73.709
        & +66.766 ($10.62\times$) \\

        \addlinespace[2pt]

        \multirow{2}{*}{128}
        & with
        & \textbf{19.826}
        & \textbf{5.969}
        & \textbf{3.216}
        & \textbf{2.917}
        & \textbf{2.425}
        & -- \\
        & without
        & 73.837 & 73.507 & 73.299 & 73.232 & 73.182
        & +70.757 ($30.18\times$) \\

        \addlinespace[2pt]

        \multirow{2}{*}{256}
        & with
        & \textbf{16.285}
        & \textbf{10.225}
        & \textbf{7.832}
        & \textbf{1.896}
        & \textbf{1.645}
        & -- \\
        & without
        & 73.271 & 72.795 & 72.533 & 72.488 & 72.452
        & +70.807 ($44.04\times$) \\

        \midrule

        \multirow{2}{*}{Mean}
        & with
        & \textbf{20.531}
        & \textbf{11.687}
        & \textbf{9.259}
        & \textbf{6.190}
        & \textbf{5.480}
        & -- \\
        & without
        & 74.624 & 73.915 & 73.708 & 73.557 & 73.460
        & +67.980 ($13.41\times$) \\

        \bottomrule
    \end{tabular*}
\end{table*}

As shown in Table~\ref{tab:physics_encoding_ablation}, removing conditioning from the physics-conditioned autoregressive NNQS consistently increases the mean absolute energy error across all model widths and optimization checkpoints. At 50K steps, the error increases from 10.908 to 74.498~mHa for the width-32 model and from 1.645 to 72.452~mHa for the width-256 model, corresponding to degradation factors of $6.83\times$ and $44.04\times$, respectively. Averaged across the four model widths, the final error increases from 5.480 to 73.460~mHa, representing a $13.41\times$ degradation. Notably, the ablated models remain at approximately 72--75~mHa throughout optimization, whereas the physics-conditioned autoregressive NNQSs exhibit substantial and sustained error reduction.
This contrast highlights the importance of the combined physics-conditioning pathway in the shared multi-molecule setting. By encoding molecular geometry and Hamiltonian information into a molecule-specific condition and injecting it into the neural ansatz, the model can adapt its predicted wave-function amplitudes and phases to different molecular Hamiltonians while retaining shared network parameters. Without this conditioning, the occupation sequence alone is insufficient to distinguish the molecular problem, substantially limiting effective optimization across multiple molecules.

\FloatBarrier

\clearpage
\setcounter{figure}{0}
\renewcommand{\figurename}{Extended Data Figure}
\renewcommand{\thefigure}{\arabic{figure}}
\setcounter{table}{0}
\renewcommand{\tablename}{Extended Data Table}
\renewcommand{\thetable}{\arabic{table}}



\end{document}